\documentclass[a4paper,11pt]{article}
\usepackage{pos}
\usepackage{amsmath}
\usepackage{bm}
\usepackage{feynmp-auto}
\usepackage{subfig}
\usepackage[makeroom]{cancel}
 
\title{\boldmath$B$ meson semileptonic decays from lattice QCD}
\ShortTitle{\boldmath$B$ meson decays from LQCD}

\author*[a,b]{Alejandro Vaquero Avil\'es-Casco}

\affiliation[a]{Departmento de Física Teórica, Universidad de Zaragoza, \\
                C/ Pedro Cerbuna 12, CP 50009, Zaragoza, Spain}
\affiliation[b]{CAPA (Center for Astroparticles and High Energy Physics), \\
                C/ Pedro Cerbuna 12, CP 50009, Zaragoza, Spain}
\emailAdd{alexv@unizar.es}
 
\abstract{$B$ processes are a rich source of potential anomalies that could lead to the discovery of BSM physics.
          The long-standing tension between the inclusive and the exclusive determinations of the CKM matrix elements $|V_{xb}|$, or the current tensions in the $R(D)$--$R(D^\ast)$ are some examples of active areas of research
          where we might find signals of new physics.
          Heavy-to-heavy $B$ semileptonic decays, $B_{(s)}\to D_{(s)}^{(\ast)}\ell\nu$, and in particular, decays with a vector product ($D^\ast_{(s)}$) are especially interesting from an experimental point of view,
          but experiment and theory must walk together in order to reach conclusions in the intensity frontier.
          In this review I talk about the current status of the lattice-QCD calculations of the $B\to D^\ast\ell\nu$ form factors at non-zero recoil, I discuss the implications they have for the determination of $B$ anomalies,
          and finally I give some hints of what we can expect from future calculations.}

\FullConference{9th Symposium on Prospects in the Physics of Discrete Symmetries (DISCRETE2024)\\
 2–6 Dec 2024\\
Ljubljana, Slovenia\\}

\begin{document}
\maketitle

\section{The Motivation}
\subsection{Indirect searches for new physics in the intensity frontier}
Although the Standard Model (SM) can predict physical phenomena with great accuracy and precision, we have found indisputable signals that there is physics Beyond the SM (BSM).
But being the SM a renormalizable theory, we do not know at which energy scale we are going to find new physics.
The LHC took us up to the $\sim O(10)$ TeV scale, and still we failed to find new particules or interactions.
We have strong indications that quantum gravity effects become relevant at the Planck scale, and certainly these are new physics not described by the SM.
But the Planck scale lies 13 orders of magnitude beyond the current reach of the LHC, which makes the direct exploration not feasible.

In this context, the intensity frontier and indirect searches play a critical role to find new physics.
The exploration of discrepancies between the theoretical predictions of the SM and the experimental measurements can open a window into which the secrets of new physics are revealed.

\subsection{The flavor sector of the SM}
\begin{figure}[h]                                                                                                                                                                  
  \includegraphics[width=\textwidth,angle=0]{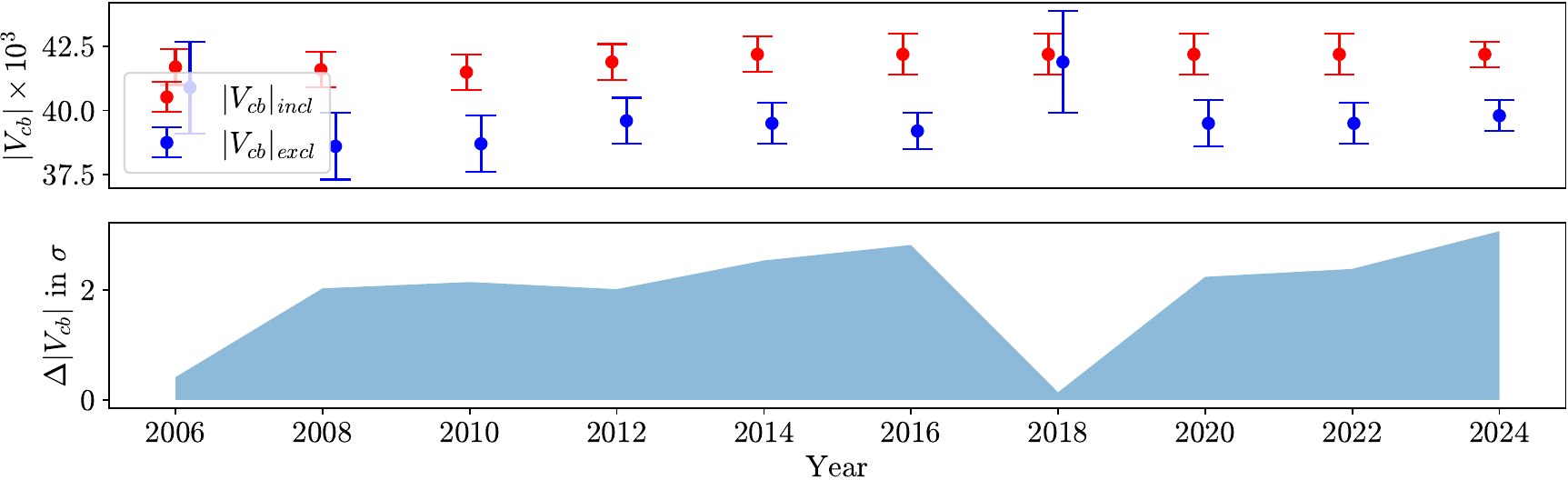}
  \caption{Difference in $\sigma$s between the inclusive and the exclusive determinations of $|V_{cb}|$ plotted against the year.
           All values of $|V_{cb}|$ come from the PDG~\cite{ParticleDataGroup:2006fqo,ParticleDataGroup:2008zun,ParticleDataGroup:2010dbb,ParticleDataGroup:2012pjm,
           ParticleDataGroup:2014cgo,ParticleDataGroup:2016lqr,ParticleDataGroup:2018ovx,ParticleDataGroup:2020ssz,ParticleDataGroup:2022pth,ParticleDataGroup:2024cfk}}
  \label{VcbDiff}
\end{figure}

From the perspective of indirect searches, many interesting processes are related to the flavor structure of the SM.
The CKM matrix elements are involved in several anomalies hinting at new physics.
In particular, $V_{cb}$ is arguably one of the most controversial ones, due to a long-standing discrepancy between its inclusive and exclusive determinations.
The discrepancy is not very large, but it certainly deserves further research.
A particularity of $V_{cb}$ is the fact that it has annoyingly kept this status for the last 17 years.
A summary is shown in Fig.~\ref{VcbDiff}.

\begin{figure}[h]
  \begin{center}                                                                                                                                                                  
  \includegraphics[width=0.6\textwidth,angle=0]{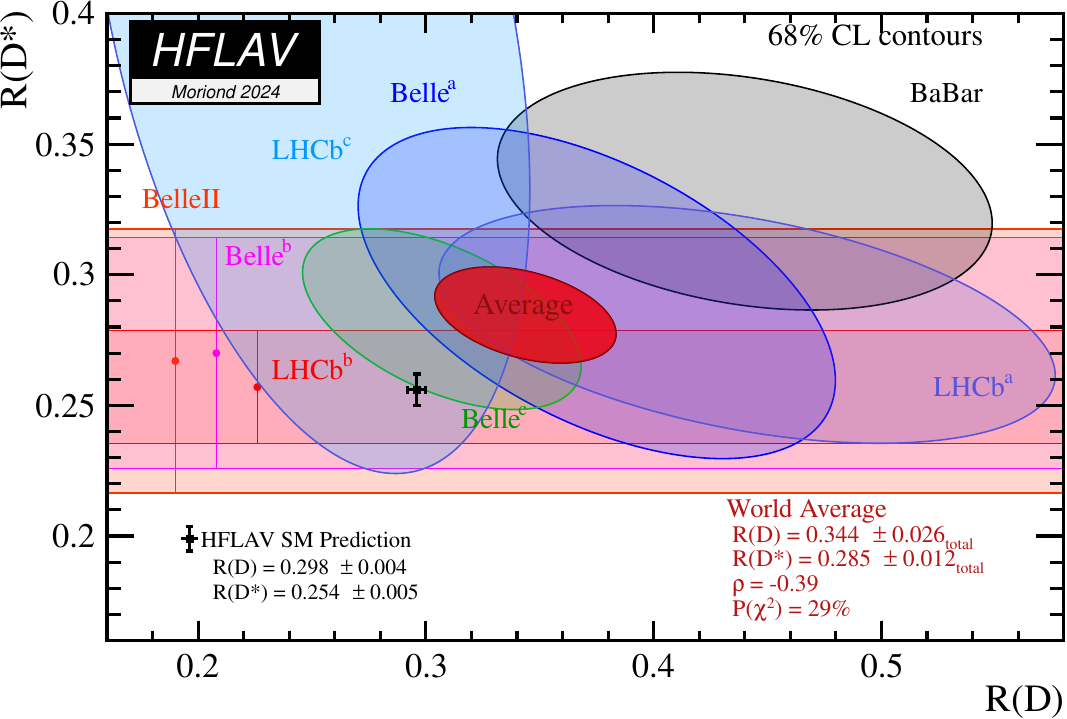}
  \caption{Tensions between theory and experiment in the $R(D)$ vs $R(D^\ast)$ plane.}
  \label{RDvsRDst}
  \end{center}                                                                                                                                                                  
\end{figure}

Another interesting set of observables come from the so called Lepton Flavor Universality (LFU) ratios, defined, for the $B\to D^{(\ast)}\ell\nu$ decays, as
\begin{equation}
R(D^{(\ast)}) = \frac{\mathcal{B}(B\to D^{(\ast)}\tau\nu)}{\mathcal{B}(B\to D^{(\ast)}\ell\nu)},\quad\ell = e,\mu,
\label{RDst}
\end{equation}
for the particular case of a $B\to D^{(\ast)}\ell\nu$ decay.
Many hadronic uncertainties are canceled in these ratios, making them excellent candidates to signal new physics.
Focusing again on the $B\to D^{(\ast)}\ell\nu$ processes, the current tension between experimental measurements and theoretical predictions in the $R(D)$-$R(D^\ast)$ plane combine to $\approx 3.3\sigma$ (see Fig.~\ref{RDvsRDst}).
The Belle II and LHCb experiments are working hard to reduce the experimental uncertainty.
From the theory side it is extremely important that we mirror these efforts.

\section{The Theory}
\subsection{Description of exclusive $B$ decays in the SM}
The differential decay rate of the $B\to D^{(\ast)}\ell\nu$ process can be written as
\begin{equation}
\underbrace{\frac{d\Gamma}{dw}(B\to D^{(\ast)}\ell\nu)}_{\textrm{Experiment}} = \left[\underbrace{K^{D^{(\ast)}}_1(w)}_{\textrm{Known factors}} \times \underbrace{\left|F(w)\right|^2}_{\textrm{Theory}} 
+ \underbrace{K^{D^{(\ast)}}_2(w,m_\ell)}_{\textrm{Known factors}} \times \underbrace{\left|F_2(w)\right|^2}_{\textrm{Theory}}\right] \times \left|V_{cb}\right|^2,
\label{BtoDstDR}
\end{equation}
where $w=v_{D^{(\ast)}}\cdot v_B$, known os the recoil parameter, is the product of the four-velocities of the $D^{(\ast)}$ and the $B$ meson; $K^{D^{(\ast)}}_{1,2}$ comprise
known factors (kinematic factors, constants, etc) defined in the theory; and $F=\mathcal{G},\mathcal{F}$ is the decay amplitude for the $B\to D^{(\ast)}\ell\nu$ decay that must be calculated from the theory.
The second term is proportional to the known function $K^{D^{(\ast)}}|F_2| \propto m^2_\ell$, so it only contributes meaningfully for the case of the $\tau$.
Equation~\eqref{BtoDstDR} can be used as the basis of an exclusive determination of $|V_{cb}|$, but it requires knowledge of the differential decay rate in the lhs, as well as the decay amplitudes and form factors involved in the rhs.
It is the duty of experimentalist to measure the former, whereas the latter is computed by theorist (mainly) using lattice QCD.

Not surprisingly, theorist and experimentalist face very different challenges: for lattice practitioners it is much easier to calculate the two form factors of the $B\to D\ell\nu$, because it just involves two form factors.
A pseudoscalar to vector decay, like $B\to D^\ast\ell\nu$, requires the calculation of four form factors, and some of them must be extracted from delicate cancelations between terms, and feature large errors.
That is why the first unquenched calculation of the $B\to D^\ast\ell\nu$ form factors away from the zero recoil point was published only a few years ago~\cite{FermilabLattice:2021cdg}.
Moreover, lattice calculations also yield smaller errors when the recoil parameter is small; the large recoil region is not easily accessible.

The experimental situation is completely opposite: the $B\to D^\ast\ell\nu$ decay enjoys a wonderful signal-to-noise ratio, and it is considered a gold plated decay, whereas the $B\to D\ell\nu$ channel suffers from a large background, coming --among others- from the large unwanted production of $D^\ast$ mesons.
Moreover, experimental measurements at small recoil, the region where lattice calculations excel, are less precise due to low statistics: the factors $K_{D^\ast}\propto (w^2-1)^{1/2}$ and $K_{D^\ast}\propto (w^2-1)^{3/2}$ kills the phase space as $w\to 1$, strongly suppressing the signal.
Given these particularities, the combination of lattice and experimental data to extract $|V_{cb}|$ from Eq.~\eqref{BtoDstDR} is not a trivial task.
 
\subsection{A reason why LFU ratios are so interesting}
The LFU ratios can be theoretically computed by integrating Eq.~\eqref{BtoDstDR} for the whole recoil range to obtain the total branching fraction,
\begin{equation}
R(D^{(\ast)}) = \frac{\int_1^{w_{\textrm{Max},\tau}} dw\, \left[K_1^{D^{(\ast)}}(w)\left|F(w)\right|^2 + K_2^{D^{(\ast)}}(w,m_\tau)\left|F_2(w)\right|^2\right] \times \xcancel{\left|V_{cb}\right|^2}}
                     {\int_1^{w_{\textrm{Max}}} dw\,K_1^{D^{(\ast)}}(w)\left|F(w)\right|^2\times \xcancel{\left|V_{cb}\right|^2}}
\end{equation}
where $|V_{cb}|$ cancels out. Hence, a pure theoretical calculation is possible.
On the other hand, total branching fractions can also be measured in experiment, opening the possibility for a direct comparison between a lattice and an experimental result.

\subsection{$B$ decays on the lattice: Description and challenges}
The calculation of the decay amplitudes $\mathcal{F}$ and $\mathcal{G}$ requires the usage of non-perturbative techniques, being lattice QCD the most successfult among those.
Lattice calculations of the form factors of the gold-plated process $B\to D^\ast\ell\nu$ have become a primary topic of discussion in any flavor physics review related to BSM searches.
The SM form factors for this decay can be obtained from the following expressions:
\begin{align}
\frac{\left\langle D^\ast(p_{D^\ast},\epsilon^\nu)\right|\mathcal{V}^\mu\left|\bar{B}(p_B)\right\rangle}{2\sqrt{M_B\,M_{D^\ast}}} =&\quad
\frac{1}{2}\epsilon^{\nu *}\varepsilon^{\mu\nu}_{\,\,\rho\sigma} v_B^\rho v_{D^\ast}^\sigma \bm{h_V}(w), \\
\frac{\left\langle D^\ast(p_{D^\ast},\epsilon^\nu)\right|\mathcal{A}^\mu\left|\bar{B}(p_B)\right\rangle}{2\sqrt{m_B\,m_{D^\ast}}} =&\quad
\frac{i}{2}\epsilon^{\nu *}\left[g^{\mu\nu}\left(1+w\right)\bm{h_{A_1}}(w) - v_B^\nu\left(v_B^\mu \bm{h_{A_2}}(w) + v_{D^\ast}^\mu \bm{h_{A_3}}(w)\right)\right],
\end{align}
where $\mathcal{V}^\mu$ and $\mathcal{A}^\mu$ are the vector and axial currents in the continuum, $M_{B,D^\ast}$ are the rest masses of the mesons and
$p_{B,D^\ast}$ their momenta, $\varepsilon^{\mu\nu}_{\,\,\rho\sigma}$ is the fully antisymmetric tensor, $\epsilon^{\nu *}$ represents the polarization of the vector meson, and the $h_X(w)$, which have been highlighted in bold, are the different form factors.

The decay amplitude $\mathcal{F}$ and $F_2$ for the case of the $\tau$ can be easily reconstructed from the $h_X$, and then be used in Eq.~\eqref{BtoDstDR} to perform an exclusive determination of $|V_{cb}|$ by including experimental data.
Inconsistent results might be a signal that new physics are at play.

Lattice QCD calculations require the discretization of the gauge and the fermionic actions, in order to make them treatable for supercomputers.
The process can be done in many different ways, and lead to different versions of the actions, each one with different properties at finite lattice spacing $a > 0$, but converging to the same result in the continuum.
Choosing the right regularization for the fermionic action is critical for calculations involving heavy quarks, because the lattice spacings currently available are not fine enough to allow for a safe simulation of the bottom quark:
discretization errors grow as $\alpha^k(am_Q)^n$, which can easily get out of control for $am_Q \gtrsim 1$.
Ideally we would seek $am_b \ll 1$, but most state-of-the-art simulations are performed at $a\approx 0.12-0.045$ fm, which translates into $1.6-4.4$ GeV$^{-1}$, leading to large discretization errors.

A possibility is to use an effective action to simulate the heavy quarks.
This approach allows to reach physical heavy quark masses, at the expense of the introduction of a matching procedure for the Effective Field Theory (EFT).
The renormalization procedure also becomes more complicated, resulting in the introduction of new systematic errors.

Another possibility, if the lattice spacings are small enough, is to calculate the form factors at unphysical values of the heavy quark mass, to extrapolate to the physical mass at the end of the calculation.
In this approach the renormalization procedure is straightforward, but the extrapolation it introduces is far from trivial, because the value of $m_b$ affects the possible kinematic range of the decay.
For that reason, masses close to the physical value are required, which in turn means small enough lattice spacings.

Overall both methods give reasonable results, although the second technique is being favored as the available computing power increases.

\section{The Calculation}
\subsection{The Fermilab-MILC calculation of the $B\to D^\ast\ell\nu$ form factors}
The difficulties outlined in the previous section have hindered the lattice QCD calculations of the $B\to D^\ast\ell\nu$ decay for many years.
It was only until a few years ago that the form factors for this process are available away from the zero recoil point, in a pioneering calculation published by the Fermilab - MILC collaboration~\cite{FermilabLattice:2021cdg}.

\begin{figure}[h]
  \begin{center}
  \includegraphics[width=0.6\textwidth,angle=0]{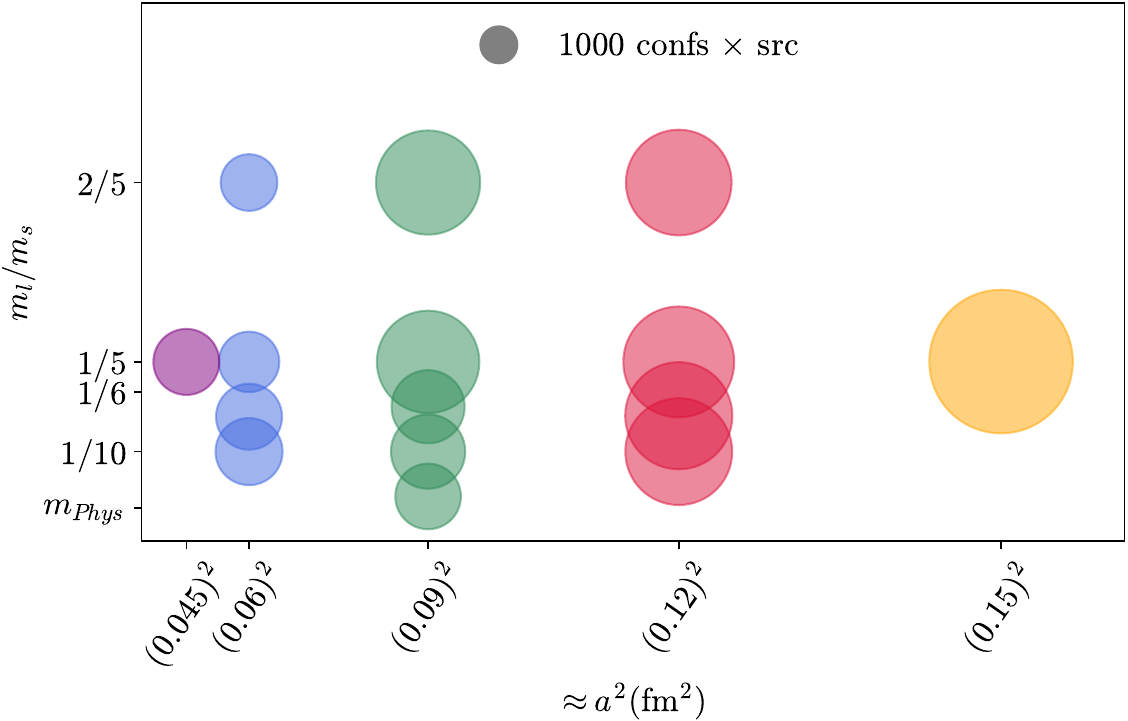}
  \caption{Ensembles employed in the Fermilab-MILC analysis. The area of each circle is proportional to the available statistics for the corresponding ensemble.}
  \label{FMEnsembles}
  \end{center}
\end{figure}

This calculation employed 15 ensembles of $N_f=2+1$ asqtad sea quarks, with the strange quark tuned to its physical mass.
The heavy quarks are simulated using an EFT, and their masses are tuned to their physical values.
The ensembles differ in the values of their light quark mass and the lattice spacing, as shown in Fig.~\ref{FMEnsembles}.
The lightest pion mass reaches $m_\pi\approx 180$ MeV, which although light, makes the $D^\ast$ meson stable.
The data obtained from the lattice correlators is used to reconstruct the form factors at finite lattice spacing and heavier-than-physical light quark masses.
Then a chiral-continuum fit is performed to take the unphysical form factors to the physical limit (zero lattice spacing and physical light quark masses).
The final result shows an excellent agreement with previous results of the $h_{A_1}$ form factor at zero recoil (see for instance Ref.~\cite{Bailey:2014tva} or Ref.~\cite{Harrison:2017fmw}).
 
\begin{figure}[h]
  \begin{center}                                                                                                                                                                  
  \includegraphics[width=0.475\textwidth,angle=0]{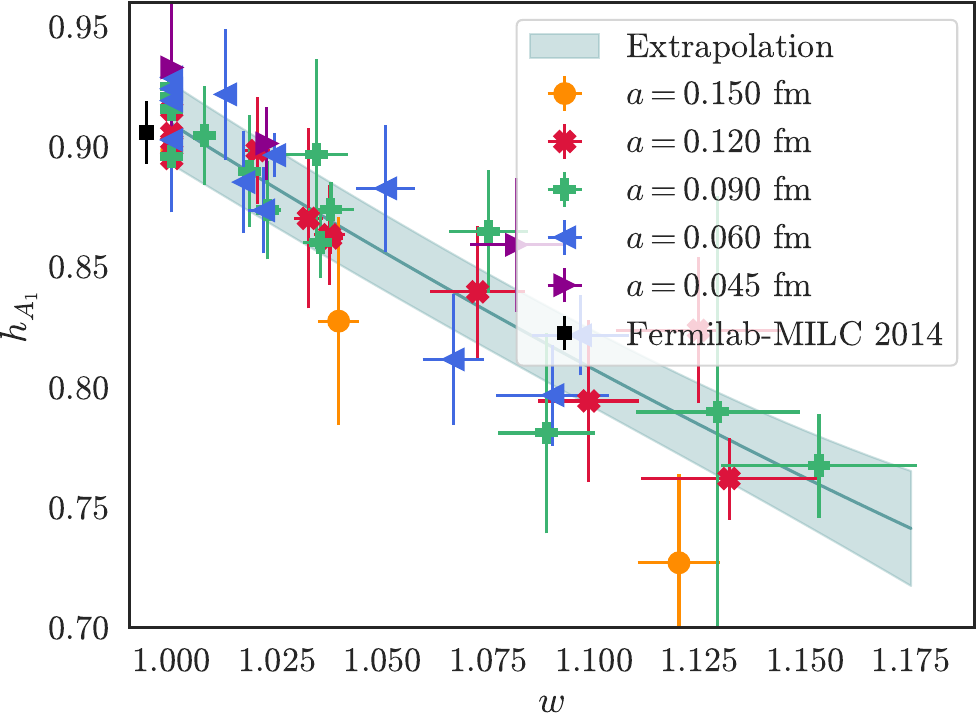} \hfill
  \includegraphics[width=0.475\textwidth,angle=0]{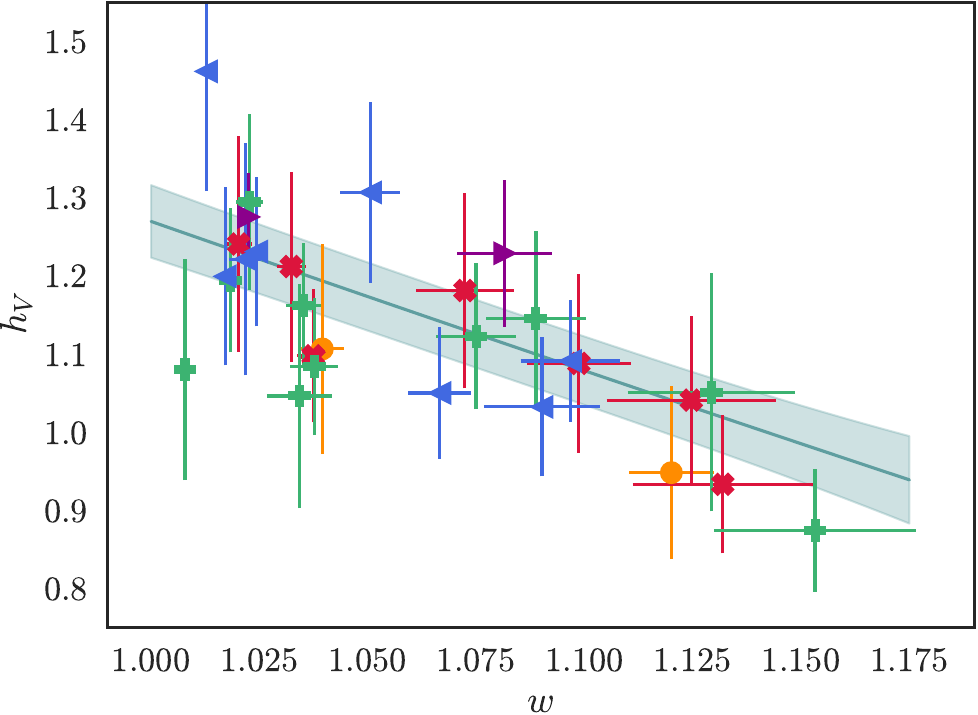}
  \caption{Chiral-continuum extrapolation of the Fermilab-MILC data for the $h_{A_1}$ (left) and the $h_V$ (right) form factors.}
  \label{FMA1V}
  \end{center}                                                                                                                                                                  
\end{figure}

\begin{figure}[h]
  \begin{center}                                                                                                                                                                  
  \includegraphics[width=0.475\textwidth,angle=0]{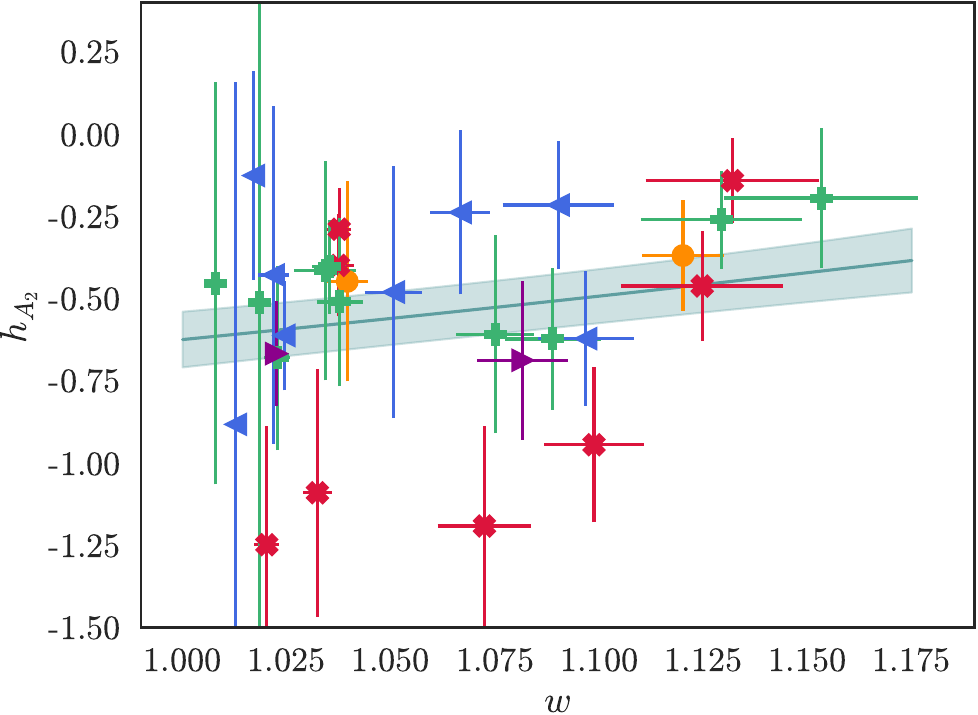} \hfill
  \includegraphics[width=0.475\textwidth,angle=0]{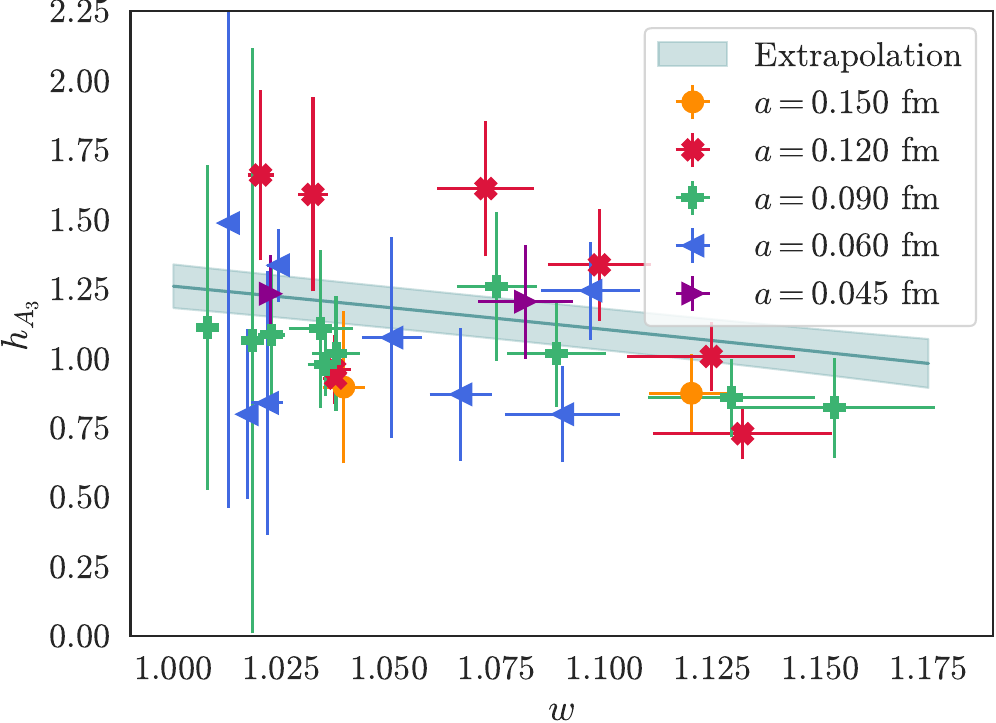}
  \caption{Chiral-continuum extrapolation of the Fermilab-MILC data for the $h_{A_2}$ (left) and the $h_{A_3}$ (right) form factors.}
  \label{FMA23}
  \end{center}                                                                                                                                                                  
\end{figure}

\begin{figure}[h]
  \begin{center}                                                                                                                                                                  
  \includegraphics[width=0.475\textwidth,angle=0]{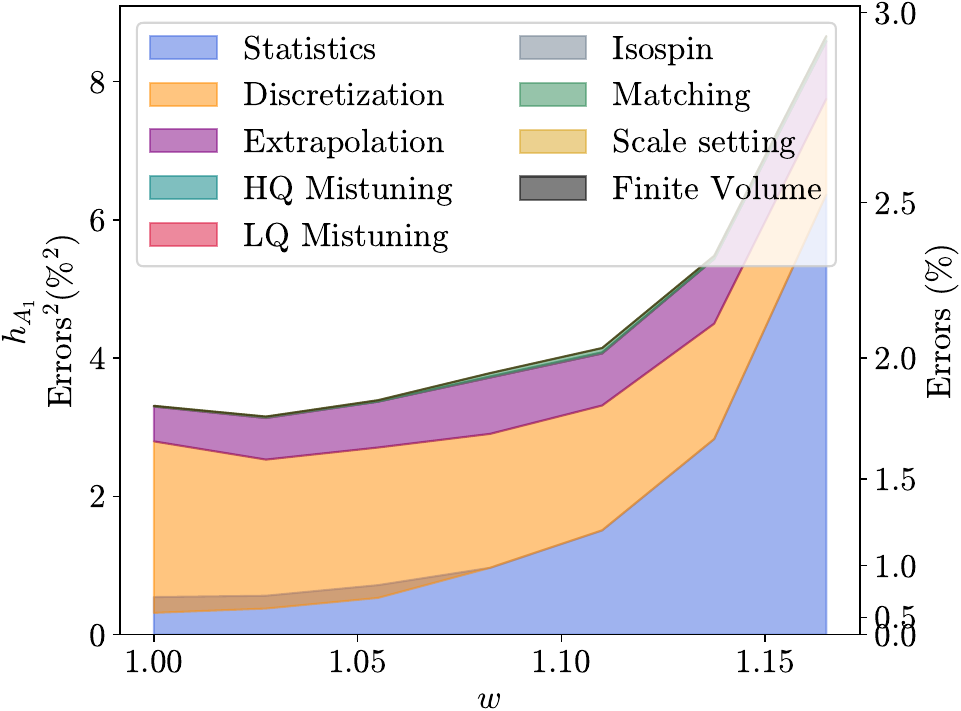} \hfill
  \includegraphics[width=0.475\textwidth,angle=0]{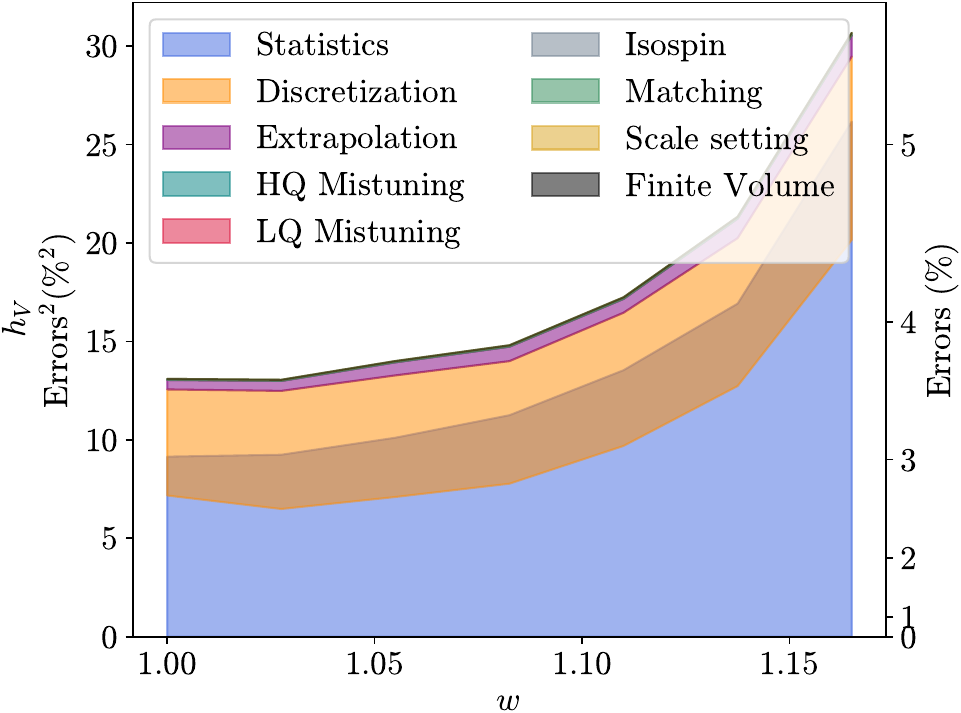}
  \caption{Complete error budget for the Fermilab-MILC $h_{A_1}$ (left) and $h_V$ (right) form factors. The $h_{A_2}$ and $h_{A_3}$ form factors display a similar distribution of errors.}
  \label{FMEBA1V}
  \end{center}                                                                                                                                                                  
\end{figure}

The continuum curves for all the form factors are shown in Figs.~\ref{FMA1V} and~\ref{FMA23}, whereas Figs.~\ref{FMEBA1V} show the error budget, obtained from a careful analysis of the systematic errors.
The main contributions come from statistics and discretization errors, especially those associated with the heavy quarks, which comes at no surprise.
 
\subsection{Beyond the lattice data: the $z$ expansion}
The exclusive determination of $|V_{cb}|$ requires a combination of experimental and theoretical data in the same recoil range, but experimental data is more precise at large recoil, whereas theoretical data is more precise precise at small recoil.
The usage of parametrizations for the form factors can bridge this gap, as well as provide a theoretical ansatz for the fits to form factor data, imposing bounds on the shapes of the different form factors.
The most interesting parametrizations are based on the connection between the pair-production process and the semileptonic decay process.
The difference between both processes is just a rotation, which can be achieved mathematically by analytically continuating a pair-production calculation of the form factors into values of $q^2$ that are only available for the semileptonic decay process.
The analytical continuation is straightforwardly performed if we use a new variable, $z$, that maps the $q^2$ corresponding to the pair-production region to the unit circle, and those corresponding to the semileptonic region onto the real axis.
Then we can write a generic form factor as,
\begin{equation}
f = \frac{1}{\phi(z)B(z)}\sum_j a_j z^j
\end{equation}
where the outer functions $\phi(z)$ are calculated from the dispersion relation, the inner functions $B(z)$ or Blaschke factors take into account the contributing poles with poorly known residues, and the coefficients $a_j$ are bounded by the \emph{weak unitarity constraints}, $\sum a_j^2 < 1$.
These are the fundamentals of the Boyd-Grinstein-Lebed (BGL) parametrization~\cite{Boyd:1995sq,Boyd:1995cf,Boyd:1997kz}, which is based only on very broad assumptions, like analyticity of the form factors or unitarity of the theory, and thus it is model-independent.

Most earlier works were based on the Caprini-Lellouch-Neubert (CLN) prametrization~\cite{Caprini:1997mu}, which offered some improvements over BGL at the cost of using more aggressive assumptions that might increase the systematic uncertainties.
At the current level of precision we have reached, the usage of the CLN parametrization is heavily discouraged by the community~\cite{Gambino:2020jvv}.

The BGL parametrization can be used to extend the lattice results, which are often obtained for small to moderate values of the recoil parameter, to the whole kinematic range.
A direct comparison between the shapes of the lattice-predicted decay amplitude and the one measured in experiment can be done at this stage, if a normalization procedure is introduced.
We show such a comparison in the left pane of Fig.~\ref{DecayComp}, where we clearly see that the Fermilab-MILC curve consistently stays below the experimental one, although the differences are always under $2\sigma$.
A combined fit to a BGL parametrization to extract $|V_{cb}|$ makes sense only if the shapes of both curves are similar.
Using data coming from Belle~\cite{Waheed:2018djm} and BaBar~\cite{Dey:2019bgc}, the Fermilab-MILC collaboration gives a value of $|V_{cb}| = 38.40(78)$ in line with previous exclusive determinations.
The quality of the different fits --as assessed with the deaugmented $\chi^2/\text{dof}$- can be checked in Table~\ref{CompResults}.
A moderately large $\chi^2/\textrm{dof}$ coming from the combined lattice QCD + experimental data is expected, due to the differences in the shapes, but perhaps more surprising is the large $\chi^2/\textrm{dof}$ obtained just with a combined experimental fit, where the only role played by the lattice is to fix the normalization to extract $|V_{cb}|$.
For all combinations showed here, the final results are compatible with the final value of $|V_{cb}|$ published by the Fermilab-MILC collaboration.
Although the numbers are consistent, further research is needed to understand the differences in shape of the decay amplitudes.
\begin{table}
  \caption{Quality of fit for the different fits attempted by the Fermilab-MILC collaboration. The Belle + BaBar column uses lattice QCD data only for renormalization.}
  \label{CompResults}                                                                                                                
  \begin{tabular}{rccccc}
    \hline                                                                                                                          
                    & {Lattice QCD} & {Belle + BaBar} & {Lattice + BaBar} & {Lattice + Belle} & {Lattice + both} \\
    \hline                                                                                                                          
$\chi^2/\text{dof}$ &    {0.63/1}   &    {104/76}     &     {8.50/4}      &     {111/79}      &     {126/84}     \\
    \hline
  \end{tabular}
\end{table}
Given the form factors in the whole recoil range, an $R(D^\ast)$ calculation is straightforward, resulting in a pure, theoretical, lattice QCD determination $R(D^\ast)_{\textrm{Lat}} = 0.265(13)$.
The still-large errors of this result prevent us from extracting meaningful conclusions.
The right pane of Fig.~\ref{DecayComp} shows a comparison of the different $R(D^\ast)$ results.
 
\section{The Mess}
\subsection{One martini is all right, two are too many, and three are not enough}
The Fermilab-MILC calculation was a huge step forward, and it prompted other collaborations to speed-up their efforts to present their own calculations.
As a result, we jumped from zero to hero in the span of three years, with two more additional calculations published by JLQCD and HPQCD.

The JLQCD calculation~\cite{Aoki:2023qpa} uses the domain-wall action for the quarks. This action is very expensive, but it has excellent properties, especially for light fermions.
On the other hand, it does not perform better than other actions for heavy quarks, which is the main problem we have to solve in this kind of computations.
The calculation relies on extrapolations to reach the physical bottom quark mass, and uses ensembles with three different lattice spacings, being $a\approx 0.044$ fm the smallest of them all, a fairly competitive value.

The $|V_{cb}|$ value obtained from a combined fit of Belle and JLQCD data, $|V_{cb}| = 39.19(90)\times 10^{-3}$, is in agreement with the Fermilab-MILC value, as well as with previous exclusive results, but their value of $R(D^\ast) = 0.252(22)$ also features large errors.
Moreover, the form factors resulting from this calculation align very well with the experimental expectations, a feature that raised eyebrows.
Why was the Fermilab-MILC decay amplitude systematically below the experimental result, when JLQCD result agrees so nicely?
Maybe there are unaccounted systematics in the Fermilab-MILC calculationm¡, perhaps related to the heavy quark regularization?
Then the HPQCD calculation entered the scene.

\begin{figure}[h]
  \begin{center}                                                                                                                                                                  
  \includegraphics[width=0.45\textwidth,angle=0]{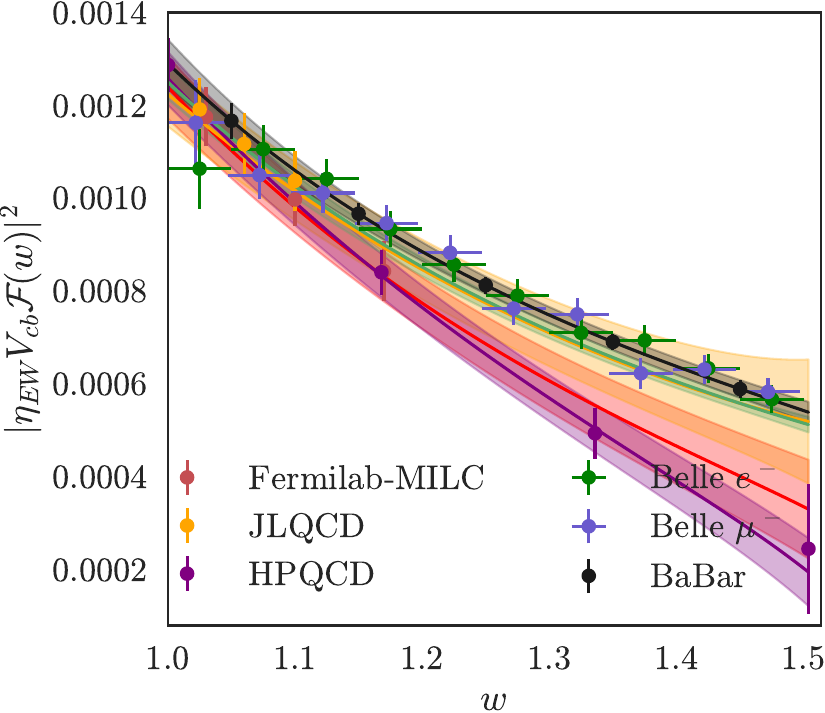} \hfill
  \includegraphics[width=0.45\textwidth,angle=0]{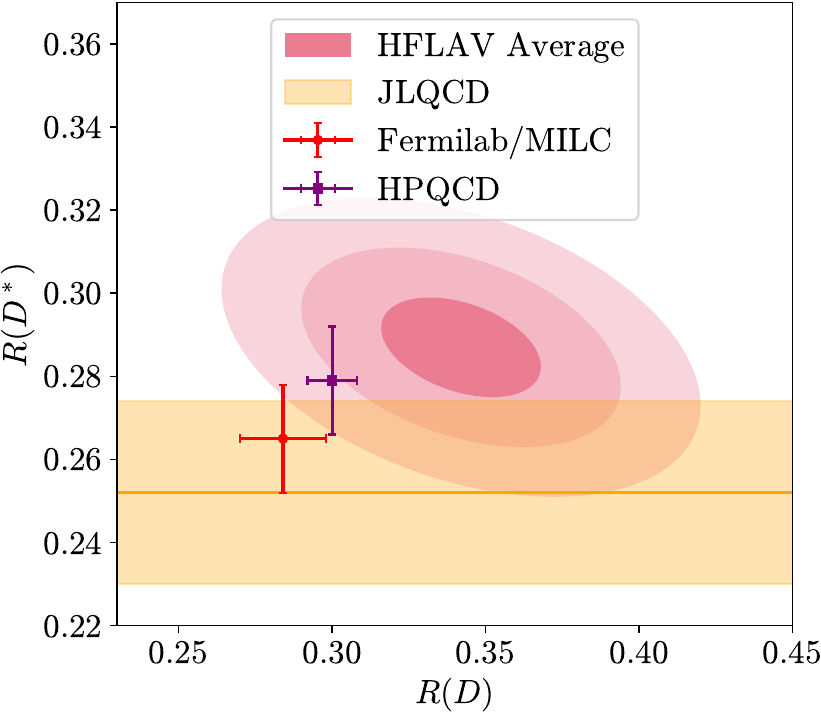}
  \caption{\textbf{Left:} Comparison of the shapes of the decay amplitude for the different calculations and experiments. \textbf{Right:} Comparison of the different values of $R(D^\ast)$ provided by lattice QCD calculations with the current HFLAV average~\cite{Banerjee:2024znd}.}
  \label{DecayComp}
  \end{center}                                                                                                                                                                  
\end{figure}

The HPQCD calculation~\cite{Harrison:2023dzh} uses the HISQ regularization for the quarks, an excellent choice given how well it performs for heavy quarks~\cite{Follana:2006rc}.
As a result, competitive heavy quark masses can be achieved with larger lattice spacings, which in turn reduces the errors in the extrapolation of the lattice results to the physical bottom quark mass.
Again, the value of $|V_{cb}| = 39.31(74)\times 10^{-3}$ is in line with previous determinations, and the value of $R(D^\ast) = 0.279(13)$ is not conclusive.
All these good features of the calculation do not translate into a better agreement between the experimental and the theoretical shape of the decay amplitude.
In fact, the shape predicted by HPQCD moves even further from the experimental value than the one shown by Fermilab-MILC.
A comparison of the three calculations is shown in the two panes of Fig.~\ref{DecayComp}.

\subsection{Lattice calculations understand each other}
However confusing the situation might seem, the lattice data, when taken away from experimental prejudices, paints a very consistent picture.
A combined BGL fit of lattice QCD data coming from three different collaborations can be used as a benchmark to measure the agreement between the different results.
A summary of such fits is given in Table~\ref{CombLat}.
These are BGL quadratic fits of the form factors, including priors $0(1)$ for each one of the coefficients.
We can compare the level of statistical agreement of the lattice results with the level of agreement reached between BaBar and Belle, whose joint fit results in a $p$-value $\approx 0.04$.
\begin{table}
  \caption{Resulting $p$-values of the different combined fits of lattice QCD data.}
  \label{CombLat}
  \begin{center}
  \begin{tabular}{rcccc}
    \hline                                                                                                                          
    & {MILC + JLQCD} & {MILC + HPQCD} & {JLQCD + HPQCD} & {MILC + JLQCD + HPQCD} \\
    \hline                                                                                                                          
$p$ &     {0.40}     &     {0.44}     &      {0.73}     &         {0.56}         \\
    \hline
  \end{tabular}
  \end{center}
\end{table}
Hence, one can understand that the different lattice results are just showing a healthy --although somewhat larger than desired- dispersion, due to the combination of statistical and systematic uncertainties coming from different approaches.
A combined fit of all lattice QCD data can be used to obtain a very competitive value of $R(D^\ast)$, coming just from theoretical considerations,
$$R(D^\ast) = 0.2667(57),$$
which, although not conclusive, shows a clear preference for the current theoretical estimates coming from other calculations.

The conclusion is that the lattice results are consistent, although the precision achieved is certainly improvable.

\section{The Future}
Even though the lattice QCD results are consistent among collaborations, that does not mean that they are satisfactory.
Given the enormous experimental efforts devoted to reduce the uncertainties in the differential decay rate measurements for this process, it would be unwise to not pursue a parallel theoretical effort to improve the form factors.

The Fermilab-MILC collaboration is working on two different calculations, aimed at improving existing results in several heavy-to-heavy decay processes.
The first calculation introduces the HISQ action for the light quarks, as well as several ensembles with physical light quark masses.
In addition, this calculation comprises four channels, the whole $B_{(s)}\to D^{(\ast)}_{(s)}\ell\nu$ decay family, so correlated results in the $R(D_{(s)})$ vs $R(D^\ast_{(s)}$ plane can be obtained.
But the calculation is even more ambitious: a parallel computation of the $B\to \pi\ell\nu$ and $B_s\to K\ell\nu$ form factors is being done, using the same setup.
This means we can produce a correlated plot in the $|V_{ub}|$ vs $|V_{cb}|$ plane.
We can expect that this calculation will bring a sizable reduction of uncertainties, but certainly it does not address the elephant in the room, which is the treatment of the heavy-quarks.
For this reason, there is another long-term effort, that is expected to reduce the errors in the form factors by one order of magnitude.

\begin{figure}[h]
  \begin{center}                                                                                                                                                                  
  \includegraphics[width=0.475\textwidth,angle=0]{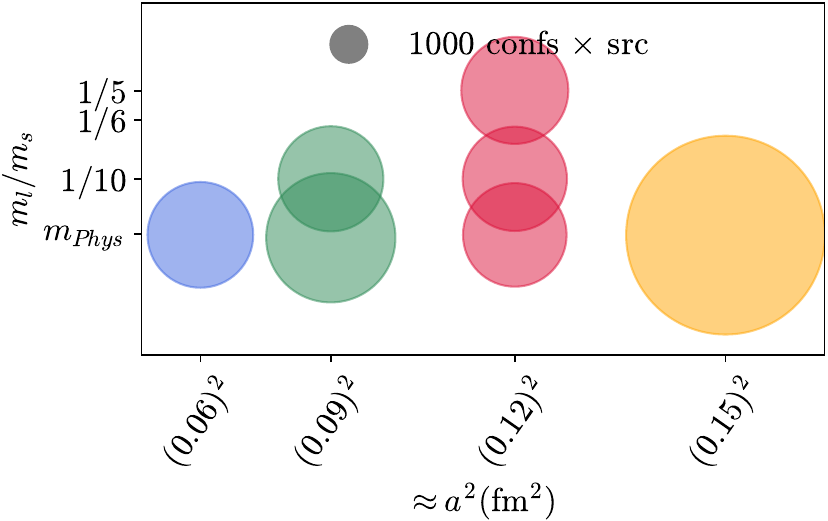} \hfill
  \includegraphics[width=0.475\textwidth,angle=0]{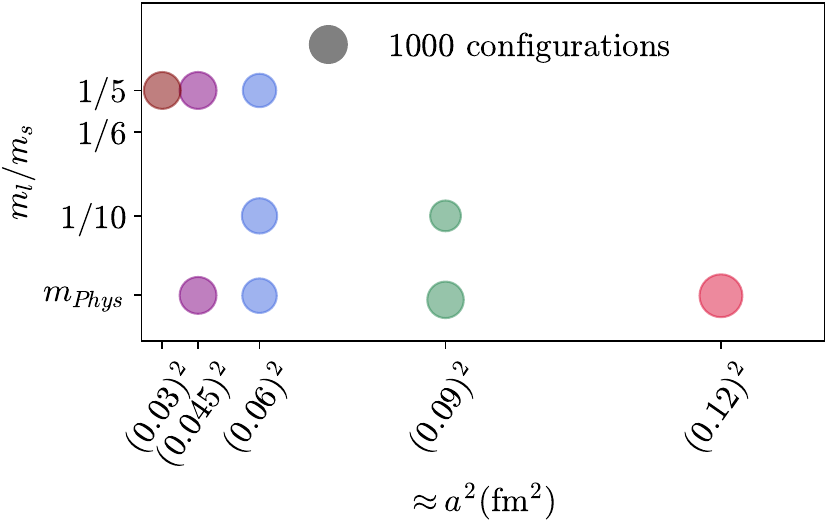}
  \caption{Complete error budget for the Fermilab-MILC $h_{A_2}$ (left) and $h_{A_3}$ (right) form factors.}
  \label{FutEns}
  \end{center}                                                                                                                                                                  
\end{figure}

The second calculation is still in early stages, meaning that the data must first be generated, so results are still a few years away.
The action chosen for both heavy and light quarks is HISQ, and the lightest lattice spacing available is $0.03$ fm, which means that physical bottom quark masses are reachable, virtually eliminating the main source of systematic errors.
The ensembles chosen for both calculations are shown in Fig.~\ref{FutEns}
 
\section{Conclusions}
The $B\to D^\ast\ell\nu$ process currently plays an important role in BSM physics searches, through the exclusive determinations of $|V_{cb}|$ and the associated LFU ratio.
Focusing on the lattice QCD result, three calculations of the form factors of this decay have appeared in the last three years.
These three calculations agree reasonably well among each other, but the level of precision attained is not enough to reach conclusions on the problems we aimed to solve.
Furthermore, there are certain tensions between some lattice predictions and experimental measurements.
Therefore, better lattice calculations are needed.

The Fermilab-MILC collaboration is committed to improve existing form factor data to match experimental efforts.
With continuous effort, and by looking at the current state of the calculations and experiments, we can reasonably expect that in the coming years we will see developments that will help answer the physical questions that have motivated these analyses.
 
\section*{Acknowledgments}
The author would like to thank T. Kaneko and J. Harrison for the preliminary materials they provided. This work was supported in part by the grants PGC2022-126078NB-C21, funded by MCIN/AEI/10.13039/501100011033 and “ERDF A way of making Europe”,
and the DGA-FSE grant 2020-E21-17R, funded by the Goivernment of Aragon and the European Union NextGenerationEU Recovery and Resilience Program on ‘Astrofísica y Física de Altas Energías CEFCA-CAPA-ITAINNOVA´.

\bibliographystyle{JHEP}
\bibliography{PoSLatDiscrete}

\end{document}